# The Rise of Certificate Transparency and Its Implications on the Internet Ecosystem

Quirin Scheitle[1], Oliver Gasser[1], Theodor Nolte[2], Johanna Amann[3], Lexi Brent[4], Georg Carle[1], Ralph Holz[4], Thomas C. Schmidt[2], Matthias Wählisch[5]

[1]TUM, [2]HAW Hamburg, [3]ICSI/Corelight/LBNL, [4]The University of Sydney, [5]FU Berlin

## ABSTRACT

In this paper, we analyze the evolution of Certificate Transparency (CT) over time and explore the implications of exposing certificate DNS names from the perspective of security and privacy. We find that certificates in CT logs have seen exponential growth. Website support for CT has also constantly increased, with now 33% of established connections supporting CT. With the increasing deployment of CT, there are also concerns of information leakage due to all certificates being visible in CT logs. To understand this threat, we introduce a CT honeypot and show that data from CT logs is being used to identify targets for scanning campaigns only minutes after certificate issuance. We present and evaluate a methodology to learn and validate new subdomains from the vast number of domains extracted from CT logged certificates.

## 1 INTRODUCTION

Certificate Transparency (CT) logs provide an append-only public ledger of TLS certificates in order to make the TLS ecosystem auditable. In April 2018, CT was made mandatory in Chrome for all newly issued certificates, for the first time offering a full view of the TLS ecosystem. This full view has pros and cons. It may increase security as owners of domain names can now verify certificates that have been issued globally for their names and thus are able to notice incorrectly issued certificates. On the negative side, CT exposes domain names in a way that eases identification of previously unknown domains and services.

In this paper, we contribute to a better understanding of CT rollout and related security and privacy implications:

**CA and CT Log Evolution (§ 2):** Using data of all CT log servers deployed, we investigate the evolution of CT logs over time and the dependency of Certificate Authorities (CAs) on CT log operators.

**Server CT Deployment (§ 3):** Using passive and active measurements, we quantify the evolution of CT adoption among server operators and show positive effects.

**DNS Information Leakage (§ 4):** We investigate the mass leakage of Fully Qualified Domain Names (FQDNs), and use subdomain data to construct and query new FQDNs.

**Detecting Phishing Domains (§ 5):** We show that CT logs can be used to detect and study phishing domains.

**CT Honeypot (§ 6):** We introduce a CT honeypot to show that third parties monitor CT logs to initiate likely malicious scans.

We aim to fully support reproducible research [37] and publish data and code under https://mediatum.ub.tum.de/1452291

## 2 TIMELINE OF CT LOG EVOLUTION

CT aims to make CA-issued certificates transparent by publishing them to CT logs, ideally operated by independent parties. This allows to catch and attribute mis-issuances sooner. Logs are append-only and use Merkle Hash Trees, which allows to detect tampering with a log's history. For every logged certificate, the log creates a Signed Certificate Timestamp (SCT), which serves as an inclusion promise and which can be verified using the log's public key. SCTs can be sent inside a TLS extension, as part of a stapled Online Certificate Status Protocol (OCSP) response, or embedded in the certificate. To embed a SCT in a certificate, a CA must submit a so-called precertificate to a CT log. The log returns an SCT, which the CA can then embed in the final certificate.

From its beginnings as an RFC proposed by Google, Certificate Transparency has seen a strong interest on the side of Web infrastructure providers. However, at the release time of the initial experimental RFC 6962 [22] only few certificates showed up in CT logs—mainly Google and Go Daddy certificates were logged to Google repositories. Relevant counter-incentives against publishing in these logs exist, mainly related to privacy, business protection, and security, as we will detail in Sections 4 and 6.

To enforce deployment, Google, in its unique position of controlling a large portion of the browser market, evolved Chrome CT policy over time, from EV-indicator-only to all certificate types requiring diversely operated log entries [5]. Towards a universal requirement, an initial announcement [38] was made in October 2016, that as of October 2017, Google Chrome would only mark new certificates trusted if they complied with Chrome's Certificate Transparency policy. Still, use by CAs remained relatively weak

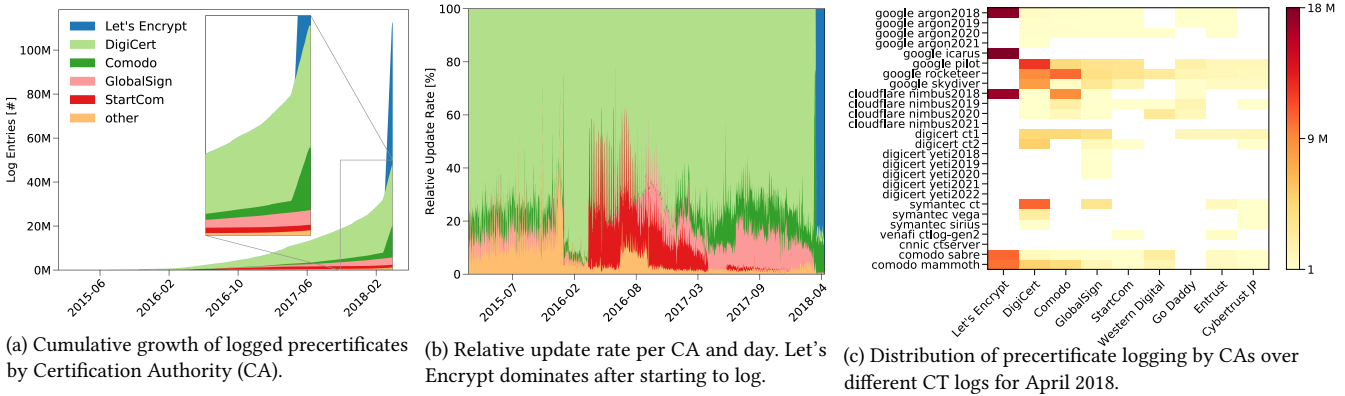

(a) Cumulative growth of logged precertificates by Certification Authority (CA).

(b) Relative update rate per CA and day. Let's Encrypt dominates after starting to log.

(c) Distribution of precertificate logging by CAs over different CT logs for April 2018.

Figure 1: Cumulative logged precertificate growth per CA, relative log rates of CAs, and distribution of logging by CA and Log.

as can be seen from Figure 1a, in which we display the cumulative growth of precertificates in all trusted CT logs over time. All data was harvested directly from the publicly available logs. Precertificates, which are untrusted issuance promises, can, unlike final certificates, only be published by Certificate Authorities themselves.

In a second attempt in April 2017, and in alignment with fixes to the initial RFC 6962, Google pushed for April 18, 2018 as the new date of strict CT policy enforcement by Chrome [29, 39]. Widely perceived, this deadline was taken seriously and deployment activities started. The top five issuing CAs (subsuming various Issuer-CNs), which accounted for 99% of the certificates in April, 2018, increased certificate publishing exponentially with pronounced final jumps starting in March, 2018 (Fig. 1a). It is interesting to follow the different time spans chosen by the CAs as visualized in Figure 1b. Over a long period, DigiCert dominated activities, followed by more irregular additions by Comodo, GlobalSign, and StartCom. In March 2018, Let's Encrypt started logging precertificates with an update rate above 2M certificates per day into few logs.

The graphs in Figure 1 show how Let's Encrypt and few other CAs dominate; the publishing behavior of a handful of CAs shapes the CT infrastructure by (un)balancing the utilization of logs. Figure 1c displays the cross-publishing relation between CAs and logs in a heatmap that is very sparsely populated. Zooming in on Let's Encrypt (left-most column) shows a worrying trend: besides Google logs, the Nimbus log is carrying the main load. This has recently even led to performance issues at Nimbus, resulting in a disqualification discussion [24]. The five big CAs publish only to a small selection of CT logs, making the ecosystem vulnerable to issues at those logs. We argue that CAs should distribute their logging load more evenly among logs and log operators.

## 3 SERVER DEPLOYMENT OF CT

This section examines actual CT server deployment in the Internet.

### 3.1 Datasets

We use two datasets. To measure the actual use of CT in the Internet, we passively monitor the Internet uplink of the University of California at Berkeley (UCB) for approximately a year. We only examine outgoing connections to prevent bias from the internal server population. We use the Bro Network Security Monitor [31]. In prior work [1], we extended Bro to support analysis and validation of Signed Certificate Timestamps (SCTs), *i.e.*, promises from logs that a certificate has been included. We extract these promises using all supported ways of transmission. For our analysis we examine traffic from 2017-04-26 to 2018-05-23. During this time we saw 26.5G TLS connections (25.6G on port 443). As our prior work [1] has shown TLS observations to yield similar results in the US, Germany, and Australia, we do not expect any geographic bias in this analysis.

To examine deployment on the Internet we perform an active Internet-wide scan of HTTPS and examine the certificates on servers. We create traffic traces and run these through Bro, resulting in the same processing pipeline for active and passive measurements. Our active scan, similar to [1, 14, 35, 36], builds on a large (≈423M) list of DNS domain names, which we resolve for A and AAAA records, conduct zmap scans on port tcp/443, and subsequently scan using a custom-built TLS scanner. We conduct weekly scans, and used a scan from May 18, 2018.

**Ethical Considerations.** For active scans, we minimize interference by following best scanning practices, such as those outlined in [11], by maintaining a blacklist and using dedicated servers with informing rDNS names, websites, and abuse contacts. We assess whether data collection can harm individuals or reveal private information as proposed by [10, 30]. Our passive data collection was cleared by UCB. Note that the data collection specifically excludes or anonymizes sensitive information, such as client IP addresses. Additionally, passively collected data never leaves institute systems. For more information about collected data, see [1].

### 3.2 CT Adoption

We take a look at how Certificate Transparency is currently adopted. We do this by examining TLS traffic at UCB and inspecting SCTs. These contain the signature of a CT log that promises the inclusion of a certificate. An SCT can be contained in a certificate, sent separately in a TLS extension, or sent in a stapled OCSP reply.

In total over our measurement period, 8.6G (32.61%) of the observed connections contained at least one SCT. 5.7G (21.40%) contained at least one SCT in the certificate, 3G (11.21%) at least one SCT in the TLS extension and 2M (>0.01%) contained at least one SCT in a stapled OCSP response. Connections where an SCT was seen via several transmission methods are relatively rare. 30.8K connections contained an SCT in both the certificate and the TLS extension; 29 in both the certificate and an stapled OCSP reply. Connections that contain the SCT in both the OCSP reply and the TLS extensions are more common: this happens in 1.5M connections. Figure 2 shows the percentage of connections that contain an SCT, split by transmission type. As we can see the number of connections containing an SCT stays relatively constant, even after Chrome enforcement started in April 2018. We assume that this picture will change in the near future with gradual certificate replacement, and given the extreme increase in logging as seen in Figure 1a. We manually examined several of the peaks in Figure 2; they were caused by large amounts of requests to *graph.facebook.com*. We are unable to determine the root-cause for this abnormality with our data. Table 1 shows the logs that we see used in traffic, split by transmission type. As we can see a small number of logs dominate.

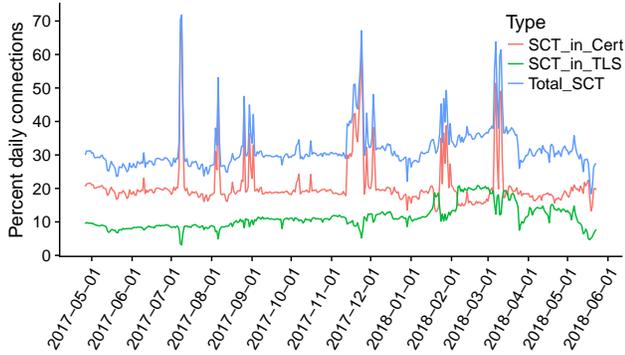

Figure 2: Percent daily connections containing an SCT (*sctconns*), split by transmission mode of embedding (*cert*) or TLS extension (*tls*). OCSP not included due to their rarity.

Table 1: Top 15 CT logs by number of observed connections

| CT Log (Chrome inclusion date) | Cert SCTs ↓ | TLS SCTs |
| --- | --- | --- |
| Google Pilot log (6/14) | 5.11G (28.69%) | 1.58G (26.03%) |
| Symantec log (9/15) | 3.28G (18.40%) | 2.44G (40.19%) |
| Google Rocketeer log (4/15) | 3.09G (17.33%) | 1.42G (23.30%) |
| DigiCert Log Server (1/15) | 1.78G (10.01%) | 9,533 (0.00%) |
| Google Skydiver log (11/16) | 1.06G (5.97%) | 54.25M (0.89%) |
| Google Aviator log (6/14) | 1.05G (5.94%) | 10,730 (0.00%) |
| Venafi log (10/15) | 994.85M (5.58%) | 148.91M (2.45%) |
| DigiCert Log Server 2 (6/17) | 671.56M (3.77%) | 12.98M (0.21%) |
| Symantec Vega log (2/16) | 661.21M (3.71%) | 1.33M (0.02%) |
| Comodo Mammoth CT log (7/17) | 78.42M (0.44%) | 225.10M (3.71%) |
| Cloudflare Nimbus2018 Log (3/18) | 8.56M (0.05%) | 104K (0.00%) |
| Google Icarus log (11/16) | 7.30M (0.04%) | 4,488 (0.00%) |
| Cloudflare Nimbus2020 Log (3/18) | 4450213 (0.02%) | 13,325 (0.00%) |
| Comodo Sabre CT log (7/17) | 2.66M (0.01%) | 120.42M (1.98%) |
| Certly.IO log (4/15) | 1.53M (0.01%) | 1 (0.00%) |

As for client support, in 17.7G (66.76%) of connections the client signals its support for the SCT extensions.

### 3.3 Server Support

We investigate server-side deployment of CT with active scans.

29.5M (68.7%) of the 42.8M unique certificates that we encounter in our scan have an embedded SCT. Furthermore, for 335.7K unique certificates the server sends a SCT in the TLS extension; for 1,214 certificates in the stapled OCSP reply. In total, 3.7M IPs serve an SCT for at least one of their hosted sites. With the use of TLS-SNI, this ≈12-fold multiplexing of certificates per IP is expected.

Looking at the SCTs contained in certificates, the picture is very different from our passive observation in Table 1: 74% of certificates contain an SCT issued by the Cloudflare Nimbus2018 Log; 71% from the Google Icarus log. The next most common log is the Google Rocketeer log (19.04%) and the Comodo Sabre CT log (12.52%). SCTs from other logs are contained in less than 10% of certificates.

This shows that characteristics of certificates generally *encountered by users* in the Internet vary strongly from those *offered across* the Internet. We presume this is caused by the high popularity of certain services.

### 3.4 Certificates with invalid embedded SCTs

Our previous work [1] revealed that SCTs are generally used correctly, except for few cases in which invalid SCTs are sent via TLS extensions, and one case with an invalid embedded SCT.

With more CAs having started to embed SCTs in their certificates, we re-evaluate this by examining our passive and active scan data. We find 16 certificates from 4 CAs that have invalid SCTs embedded. We inquired with the CAs to determine the reasons.

One certificate with an invalid embedded SCT was issued by TeliaSonera [40]. Inquiring with TeliaSonera revealed that this certificate was one of the first certificates they used to test CT. The certificate was a re-issuance of an earlier certificate, of which the SCT was included in the newer certificate. We also encounter 12 certificates issued by GlobalSign with invalid SCTs [15]. Our analysis revealed that all these certificates had Subject Alternative Names (SANs) with both DNS names and IP addresses, and that the order of entries had changed in the final certificate. We reported this to GlobalSign, who confirmed the issue and deployed a patch. Our data also contains 2 certificates with invalid SCTs issued by D-Trust [8], a German CA, who acknowledged the issue. The reason was an error in their issuance process—in some cases, the ordering of X.509 extensions differed between precertificate and final certificate, invalidating the SCT. We found one certificate issued by NetLock, a Hungarian CA, with an invalid embedded SCT [27]. Here, precertificate and final certificate contained entirely different SAN names and even issuer names. We contacted NetLock, who acknowledged the issue, re-issued the certificate, and revoked the original, but did not share a root cause.

Looking back at these issues we think that all of them can be considered birth pangs in specific and rare corner cases—CAs are still adapting to the requirements of having to embed SCTs into their certificates. When generating precertificates and final certificates, even fields without an inherent order need to be kept consistent, a requirement that CA software did not have to fulfill before.

Our disclosure of invalid SCTs to the community also fueled a discussion on whether or not CAs should log final certificates besides their precertificates. As we could provably identify issues by comparing final and precertificates, Let's Encrypt began a journey to log all final certificates [18], and the broader community also identified this as desirable [9, 19]. When Let's Encrypt initially started to log final certificates, they quickly induced performance problems in some logs [19]. This also highlights a risk of unlogged final certificates: As CT logs accept all valid certificates, a mass submission of valid unlogged final certificates could be used to overwhelm logs, which could lead to log disqualification [24].

## 4 LEAKAGE OF DNS INFORMATION

The Common Name (CN) and Subject Alternative Name (SAN) fields in certificates contain fully qualified domain names (FQDNs), often including subdomains. This makes CT a useful data source to learn about the existence of subdomains, which in turn may reveal information about the service and software for which the subdomain is used. Examples are subdomain labels such as *autodiscover* (MS Exchange); *webmail* or *smtp* for email; *api* for API access; *dev* and *staging* for development operations; *owncloud* and *citrix* for the respective products; or simply *m* for sites' mobile versions.

The leaking of DNS information was a concern about CT from the beginning: Symantec even used to operate a special log (called *Deneb*) whose explicit goal was to hide subdomains [1]. There are also efforts to standardize label redaction [17]. Subdomain enumeration is often used in the preparation of an attack and a common methodology in penetration testing. Sources like [2] even propose to query online databases such as censys.io or crt.sh when targeting particular, single domains. The bulk use of CT data has, to the best of our knowledge, not been tried yet.

The key questions we investigate are hence: how much potentially sensitive information is given away in CT, and can it be used for subdomain enumeration?

### 4.1 Data sources and processing

We describe our data sources and how we processed our data.

**Parsing DNS names.** To extract subdomains, we extract all labels under a base domain, which we define as the domain under a public suffix per Public Suffix List (PSL) [13].

**CT Data.** We extract all DNS names from CN and SAN fields of all certificates in CT logs as of 2018-04-26. Some DNS names in these fields are not valid FQDNs as defined by RFC 1035 (and later updates). We eliminate these using the Python *validators* library. Every FQDN is counted only once.

**Domain List.** For our enumeration attempts, we use the same list of registrable domain as [1]. The list includes 206M FQDNs underneath public suffixes and is mainly constructed from various large zone files, *e.g.*, *.com*, *.net*, and *.org*.

**Sonar Forward DNS.** For validation, we use the Sonar database [33] as of 2018-04-27 and parse it using the PSL. The database contains FQDNs and the result of A record DNS lookups. There are 1.3G FQDNs, of which only 1.1G have subdomains. The total count of distinct subdomain labels is 962M.

Our list and the Sonar list are relatively disjunct: 82% (37.7M) of our registrable domains (in a given public suffix) occur on the Sonar list as well (in the same public suffix). However, of the subdomain labels from our list, only 21% appear also as subdomain labels on the Sonar list (irrespective of the suffix).

### 4.2 Analysis of Subdomains

We parse the FQDNs obtained from CT as described above and count how often each subdomain label occurs across all suffixes.

Table 2: Top 20 subdomain labels (SDL) in CT-logged certificates.

| | SDL | Count | | SDL | Count | | SDL | Count |
|---|---|---|---|---|---|---|---|---|
| 1 | www | 61.1M | 8 | shop | 303k | 15 | secure | 176k |
| 2 | mail | 14.4M | 9 | **whm** | 280k | 16 | admin | 158k |
| 3 | **webdisk** | 8.7M | 10 | dev | 256k | 17 | mobile | 156k |
| 4 | webmail | 8.6M | 11 | remote | 253k | 18 | server | 146k |
| 5 | **cpanel** | 8.2M | 12 | test | 249k | 19 | cloud | 141k |
| 6 | autodiscover | 3.6M | 13 | api | 239k | 20 | smtp | 140k |
| 7 | m | 310k | 14 | blog | 235k | | | |

Not unexpectedly, this is an extreme distribution: very few subdomain labels account for by far the most occurrences. The top subdomain label, *www* accounts for 95% of subdomains, and the top 10 subdomain labels for 99% of all occurrences. The top 20 subdomain labels are shown in Table 2.

Possibly of note are *webdisk*, *cpanel* and *whm*, which all point at the existence of management interfaces, and could be interesting targets for password attacks.

We also determine the most common subdomain label for each public suffix, and find that, for example, *git* is the most common subdomain label for the suffix *tech*; *autoconfig* for *email*; *api* for *cloud*; *ftp* for *design*; *sip* for *gov*; and *dialin* for *gov.uk*—possibly reflecting the services most commonly deployed under those suffixes.

### 4.3 Subdomain Enumeration

Commonly, subdomain enumeration uses word lists to prepend words as subdomain labels to known registrable domains. Two popular hacking tools, *subbrute* [34] and *dnsrecon* [32], ship such wordlists. We test whether they would find the FQDNs that are logged in CT. *subbrute* comes with a list of 101k subdomain labels. Interestingly, just 16 of these occur as subdomain labels in logged certificates. Visual inspection of the list confirms our impression that most entries are unlikely to be common choices for subdomains. *dnsrecon* ships 1.9k names; just 12 appear as subdomain labels in CT. Since these tools would not have found real, existing FQDNs, we did not use them for further comparisons.

**Constructing FQDNs from CT data.** We construct FQDNs from subdomain labels in CT strategically and verify their existence next. We first determine which subdomain labels occur frequently in a public suffix, and prepend only these labels to the registrable domains in that suffix. We filter out any subdomain label that occurs less than 100k times in the entire data set. Both steps limit the total number of FQDNs we have to verify using DNS. We disregard the zones *.com*, *.net*, and *.org*: they are too generic for our purpose. For every subdomain label, we filter for the top 10 most common public suffixes in which it occurs. We finally prepend the subdomain label to those domains from our domain list [1] that fall into the 10 suffixes. This method leaves us with 210.7M new FQDNs to test.

**Verifying the Existence of FQDNs.** We use *massdns* to determine whether our new FQDNs have an A record. We need to rule out zones where queries for non-existing FQDNs would return a default A record. To this end, we create a second list of FQDNs, where we replace the subdomain label with a 16-character pseudo-random string.

We then scan for both the pseudorandom FQDN as well as the constructed one, following CNAME indirection up to 10 times. We disregard IP addresses not part of our border router's routing table as invalid. This rules out misconfigured DNS servers. It also makes our numbers lower bounds.

Table 3: Potential phishing domains identified in CT.

| Service | Count | Example |
|---|---|---|
| Apple | 63k | *appleid.apple.com-7etr6eti.gq* |
| PayPal | 58k | *paypal.com-account-security.money* |
| Microsoft | 4k | *www-hotmail-login.live* |
| Google | 1k | *accounts.google.co.am* |
| eBay | <1k | *www.ebay.co.uk.dll7.bid* |

We obtain 80.3M replies to our test DNS names, and 61.5M replies to our pseudo-random controls. This yields 18.8M cases of new FQDNs with previously unknown subdomains.

**Comparison to Sonar.** Of our 18.8M newly found FQDNs, only 1.1M were known via the Sonar list. This results in 17.7M newly constructed and discovered subdomains from CT, making it an additional source to infer new FQDNs and subdomains that do not yet occur in public lists.

## 5 DETECTING PHISHING DOMAINS

With the general move towards HTTPS on all sites, phishing sites need certificates as well. Hence, CT data should be useful to detect phishing domains. Facebook and CertSpotter even offer notification services for operators [12, 23] to receive advisories about potential phishing attempts against their users. However, their methods are not disclosed, and an independent analysis of efficacy is not available. We hence decided to explore the viability of using CT data to find phishing domains with a pilot experiment.

After removing legitimate domains, *e.g.*, subdomains of *apple.com* are considered legitimate Apple domains, we observe 63k domains including the string *appleid*, of which 42k have *com*, *ga*, *info*, *tk*, and *ml* public suffixes. The vast majority of these appear to mimic Apple ID log-in URLs, probably for phishing credentials. Using simple regular expression matching techniques and visual inspection, we further identify over 126k unique potential phishing domains across the five common services shown in Table 3. Our regular expressions match domains which include the name of the service or a subset of labels of its FQDN (e.g. *login.live* for Microsoft), and we exclude the service's legitimate domains. Of the eBay phishing domains, 28% use the *bid* and *review* public suffixes; similarly, 4% of Microsoft Live phishing domains use the *live* suffix, suggesting a link between phishing target and public suffix choice.

Additionally, we identify several subdomains imitating government taxation offices such as the Australian Taxation Office (*e.g.*, *ato.gov.au.eng-atorefund.com*), U.K. HM Revenue and Customs (*e.g.*, *hmrc.gov.uk-refund.cf*), and the U.S. Internal Revenue Service (*e.g.*, *refund.irs.gov.my-irs.com*).

Based on our initial findings, and the observation that many phishing domains include a variation of the target's FQDN, we believe CT data to be a very promising way to defend against phishing attacks, opening a new research direction. We have reported phishing domains to CERTs and affected companies.

## 6 CT HONEYPOT

In this section, we introduce *CT Honeypot*. Its purpose beyond a regular honeypot [26] is twofold: First, a better understanding whether data within CT logs is monitored by third parties to gather knowledge about potential new service endpoints. Second, a better understanding of activities when this knowledge is available. We assume the following attacker model: A malicious user observes CT logs to identify new DNS names. Based on this data, the attacker precisely targets victims, instead of performing uninformed scanning of the IP address space to find vulnerable services.

A CT honeypot consists of the following four building blocks: *(i)* unique random (sub-)domains that are hard to guess, *(ii)* leaking the existence of subdomains exclusively via Certificate Transparency logs (*i.e.*, creating CT-logged certificates), *(iii)* monitoring requests to the authoritative DNS server, and *(iv)* monitoring of communication to A/AAAA records of the subdomains.

### 6.1 Implementation and Setup

We create random 12-character subdomains, which we leak to CT by obtaining certificates. To prevent leakage by third parties and to closely monitor lookup activities, we control the authoritative name server for these DNS domain names.

To discern informed from arbitrary scanning, we deploy two methods: *(i)* For each subdomain, we create an AAAA record with a unique IPv6 address. We do not enter these IPv6 addresses into the rDNS tree to avoid discovery through rDNS walking, and do not use them in any other way. We also create A records for the subdomains, but due to the small address space, IPv4 addresses are not suited to discern informed scanning. *(ii)* We monitor whether scanners use the subdomains in subsequent connections, *e.g.*, as a TLS-SNI or in HTTP GET requests.

In 3 batches, we create 11 honeypot subdomains over 18 days. We store full packet captures from our monitors from 2018-04-12 14:00 UTC until 2018-05-15 14:00 UTC. We filter out DNS queries from the issuing CA's validation infrastructure, identified by DNS queries before CT logging and validated by our prior work [35].

### 6.2 Results

**DNS Queries.** After publication of precertificates for our subdomains in CT logs, we see the first DNS queries for corresponding domain names after 73 seconds to ≈3 minutes (see Table 4). This clearly highlights that CT logs are monitored. We can distinguish between two types of queries: Queries that occur among almost all domain names and queries that occur occasionally. We now analyze the DNS resolvers and their queries in more detail. It is important to note that after the first DNS lookup, a domain name may also be learned from sources other than CT, such as DNS threat intelligence networks like FarSight's DNSDB. However, the initial leakage still comes from CT logs.

We receive DNS queries from Google (AS 15169), 1&1 (AS 8560), Amazon (AS 16509), and DigitalOcean (AS 14061) for all 11 domains. From Deteque (AS 54054) we see requests for 9 domains and from OpenDNS (AS 36692) for 7 domains. Servers from these networks start querying at least one domain name in less than 12 minutes, except DigitalOcean that sends the first query after ≈2 hours. Deviations in times are not surprising even in automated settings: First,

Table 4: Per subdomain (A-K), we list its first CT log entry (all times UTC and 2018), the first DNS query, the time between CT log entry and DNS query, the total count of DNS queries (Q), the count of DNS querying ASes (AS), the count of unique EDNS client subnets (CS), the first 3 connecting ASes, the first HTTP(S) connection, and the HTTP(S) ASNs.

|   | CT log entry | DNS | $\Delta_t$ | Q | AS | CS | First 3 ASes | HTTP(S) | $\Delta_t$ | HTTP ASNs |
|---|---|---|---|---|---|---|---|---|---|---|
| A | 04-12 14:16:59 | 14:20:16 | 197s | 55 | 14 | 4 | ★15169, ▲8560, ■54054 | 04-12 15:33:49 | 73m | ◗14061,✤16509 |
| B | 04-12 14:18:31 | 14:19:44 | 73s | 55 | 14 | 3 | ★15169, ●44050, ▲8560 | 04-12 15:38:27 | 79m | ◗14061,✤14618 |
| C | 04-20 10:43:44 | 10:45:03 | 101s | 81 | 14 | 3 | ★15169, ■54054, ▲8560 | 05-10 06:44:44 | 19d | ◗14061,✤16509 |
| D | 04-30 13:00:28 | 13:02:08 | 96s | 36 | 10 | 2 | ★15169, ■54054, ▲8560 | 04-30 14:53:46 | 111m | ◗14061,✤16509 |
| E | 04-30 13:03:10 | 13:05:50 | 120s | 30 | 12 | 3 | ★15169, ▲8560, ✤16509 | 04-30 14:50:39 | 85m | ◗14061,✤16509 |
| F | 04-30 13:50:06 | 13:52:04 | 118s | 36 | 13 | 3 | ★15169, ▲8560, ✤16509. | 04-30 14:51:26 | 59m | ◗14061,✤16509 |
| G | 04-30 14:00:07 | 14:02:05 | 118s | 62 | 32 | 7 | ★15169, ▲8560, ●44050 | 05-10 06:26:51 | 5d | ◗14061,✤16509 |
| H | 04-30 14:10:07 | 14:12:04 | 117s | 32 | 11 | 3 | ★15169, ▲8560, ■54054 | 04-30 16:12:33 | 122m | ◗14061,✤16509 |
| I | 04-30 14:20:07 | 14:22:04 | 117s | 44 | 18 | 3 | ★15169, ▲8560, 24940 | 04-30 16:12:33 | 112m | ◗14061,✤16509 |
| J | 04-30 14:30:07 | 14:32:07 | 120s | 36 | 10 | 3 | ★15169, ▲8560, 12876 | 04-30 16:10:03 | 98m | ◗14061,✤16509 |
| K | 04-30 14:40:07 | 14:42:11 | 124s | 39 | 19 | 3 | ★15169, ▲8560, 19397 | 04-30 16:10:57 | 88m | ◗14061,✤16509 |

★Google, ▲1&1, ■Deteque, ●Petersburg Internet, ✤Amazon, 24940: Hetzner, 12876: Online, 19397: ACN, ◗Digital Ocean

time-triggered events may have a delay depending on probe load. Second, setups may either be run in a *streaming* fashion, using *e.g.*, CertStream [3], or in a batched fashion.

We also observe requests from 76 other ASes to one or two domains, as well as requests for three and four domains from two ASes each. In 99% of those cases, requests do not appear before one hour, in 62% before two hours. In contrast to the top servers above, we argue that those requests are initiated from batch jobs as opposed to near-real-time stream processing.

Now, looking at servers that poll data for more than 60% of our domains, we note that Deteque is a division of Spamhaus and offers DNS related threat intelligence. This business model indicates intrinsic interest in recent DNS data. Furthermore, DNS requests from Google's public DNS resolver include the *EDNS Client Subnet* field [7] in 169 cases. This DNS extension carries data about the network that originated the DNS query. It helps us to reveal the topological location of stub resolvers or clients which use Google's open recursive DNS resolver. We find 12 unique EDNS client subnets at size /24. The top 3 are used 115, 25, and 10 times, while the remaining 9 are only used 1-2 times.

Evaluating DNS lookups per included EDNS client subnet permits us to identify a few interesting patterns: First, stub resolvers in Hetzner (AS 29073) are using Google Public DNS service within few minutes and scan A, AAAA, MX, NS, and SOA records. Second, resolvers hosted in Quasi Networks (AS 29073) also very rapidly query A and AAAA records via Google Public DNS.

**Suspicious Connections.** Out of 4 of the 12 EDNS client subnets, 1 machine each connects to our honeypot over IPv4. 3 out of these 4 machines only connect to TCP port 443 (HTTPS). One machine, associated with a subnet recorded in 25 DNS queries, scanned 30 ports across our 2 machines, likely with malicious intent. This heavily-scanning host is located in Quasi Network (AS 29073). This Autonomous System has reincorporated in the Seychelles in 2015 and has since then been known to ignore all abuse messages [25]. We also note that across all inbound scans, no source IP address followed scanning best practices such as informative rDNS names, websites, or whois entries. This likely excludes benevolent scanners from academia or industrial research as responsible entities.

To our unique IPv6 addresses, no inbound packets arrived except those from the Let's Encrypt validation server.

**Conclusion.** The variety of clients frequently querying our domains within few minutes up to few hours indicates that several entities implement backends to monitor CT logs and react quickly to the appearance of new domain names. The correlation of DNS clients and port scanners also indicates that CT logs are misused to find potential targets for malicious connections. With the increase of IPv6 deployment, which challenges scanning per se [20], we expect more incidents in which CT logs are leveraged by attackers.

## 7  RELATED WORK

Although standardization of CT began mid of 2012, it only recently raised interest in the measurement community. Before CT was mandatory in Chrome, focus was on active scans to quantify coverage of various certificate sources [41] and to describe basic properties of logs and certificates in the CT ecosystem [16]. Then, CT as part of various HTTPS security extensions was analyzed [1]. Our work confirms that corner-cases in CA software can cause invalid CT certificates. Most recently, the performance impact of CT on HTTPS [28] and the deployment of sub-par certificates sourced from CT logs [14, 21] was measured. While the privacy implications and traceability of TLS certificates has been studied before [4, 6, 42], to the best of our knowledge, there is no detailed analysis on security and privacy aspects due to the rise of CT.

## 8  CONCLUSION

In this paper, we showed that the deployment of Certificate Transparency is progressing well but that this progress also introduces new threats. First, the bulk of certificates is logged to very few logs, creating a fragile ecosystem. Second, domain names of CT-logged certificates reveal information that might be considered confidential or private. Third, leaked domain names are actively used in Internet scanning, some of it likely malicious.

We agree that CT addresses a specific security vector, but, based on our study, are also very concerned about new attack vectors introduced by CT. We hope our results encourage work on countermeasures to protect Internet infrastructure.

**Acknowledgments:** The authors thank the anonymous reviewers and our shepherd Brian Trammell for their valuable feedback. This work was partially funded by the German Federal Ministry of Education and Research under project X-Check (grants 16KIS0528K, 16KIS0529, and 16KIS0530), by the National Science Foundation under grant numbers CNS-1528156 and ACI-1348077, and by an ECR grant of the University of Sydney. Any opinions, findings, and conclusions or recommendations expressed in this material are those of the author(s) and do not necessarily reflect the views of the German Federal Ministry of Education and Research or the NSF.

## REFERENCES

[1] Johanna Amann, Oliver Gasser, Quirin Scheitle, Lexi Brent, Georg Carle, and Ralph Holz. Mission Accomplished? HTTPS Security after DigiNotar. In *IMC*, 2017.

[2] Appsecco. The art of sub-domain enumeration. https://github.com/appsecco/the-art-of-subdomain-enumeration, 2018.

[3] Cali Dog Security. Certstream. https://certstream.calidog.io/, May 1, 2018.

[4] Frank Cangialosi, Taejoong Chung, David Choffnes, Dave Levin, Bruce M Maggs, Alan Mislove, and Christo Wilson. Measurement and analysis of private key sharing in the https ecosystem. In *SIGSAC Conference on Computer and Communications Security*. ACM, 2016.

[5] Chromium. Certificate Transparency in Chrome. https://github.com/chromium/ct-policy/blob/master/ct_policy.md, 2018.

[6] Taejoong Chung, Yabing Liu, David Choffnes, Dave Levin, Bruce MacDowell Maggs, Alan Mislove, and Christo Wilson. Measuring and applying invalid ssl certificates: the silent majority. In *IMC*. ACM, 2016.

[7] C. Contavalli, W. van der Gaast, D. Lawrence, and W. Kumari. Client Subnet in DNS Queries. RFC 7871, IETF, May 2016.

[8] D-Trust. 2 Certificates with Invalid Embedded SCT. https://misissued.com/batch/40/, 2018.

[9] Tom Delmas. M.D.S.P: Submission to ct-logs of the final certificate when there is already a pre-certificate . https://groups.google.com/d/topic/mozilla.dev.security.policy/VBnApSMUXTw/discussion, 2018.

[10] David Dittrich, Erin Kenneally, et al. The Menlo Report: Ethical Principles Guiding Information and Communication Technology Research. *US Department of Homeland Security*, 2012.

[11] Zakir Durumeric, Eric Wustrow, and J. Alex Halderman. ZMap: Fast Internet-wide Scanning and Its Security Applications. In *USENIX Security*, 2013.

[12] Facebook. Certificate Transparency Monitoring Tool. https://developers.facebook.com/docs/certificate-transparency, 2018.

[13] Mozilla Foundation. https://publicsuffix.org/, 2018.

[14] Oliver Gasser, Benjamin Hof, Max Helm, Maciej Korczynski, Ralph Holz, and Georg Carle. In Log We Trust: Revealing Poor Security Practices with Certificate Transparency Logs and Internet Measurements. In *PAM*, 2018.

[15] GlobalSign. 12 Certificates with Invalid Embedded SCT. https://misissued.com/batch/39/, 2018.

[16] Josef Gustafsson, Gustaf Overier, Martin Arlitt, and Niklas Carlsson. A First Look at the CT Landscape: Certificate Transparency Logs in Practice. In *PAM*, 2017.

[17] Kirk Hall and Rob Stradling. CT Domain Label Redaction. https://cabforum.org/pipermail/public/2017-November/012458.html, 2018.

[18] Hanno Böck. Let's Encrypt Forum: Non-Logging of Final Certifictes. https://community.letsencrypt.org/t/non-logging-of-final-certificates/58394, 2018.

[19] Jacob Hoffman-Andrews. Logging of final certificates and availability. https://groups.google.com/a/chromium.org/forum/#!topic/ct-policy/03pvlpmMpeI, 2018.

[20] Johannes Klick, Stephan Lau, Matthias Wählisch, and Volker Roth. Towards Better Internet Citizenship: Reducing the Footprint of Internet-wide Scans by Topology Aware Prefix Selection. In *IMC*, 2016.

[21] Deepak Kumar, Zhengping Wang, Matthew Hyder, Joseph Dickinson, Gabrielle Beck, David Adrian, Joshua Mason, Zakir Durumeric, J. Alex Halderman, and Michael Bailey. Tracking Certificate Misissuance in the Wild. In *IEEE S&P*, 2018.

[22] B. Laurie, A. Langley, and E. Kasper. Certificate Transparency. RFC 6962, IETF, June 2013.

[23] SSL Mate. CertSpotter. https://sslmate.com/certspotter/, 2018.

[24] Brendon McMillion. Post-Mortem: Nimbus issuing bad SCTs . https://groups.google.com/a/chromium.org/forum/#!topic/ct-policy/E88pjOZzkIM, 2018.

[25] Troy Mursch. A conversation with RIPE NCC regarding Quasi Networks LTD. https://badpackets.net/a-conversation-with-ripe-ncc-regarding-quasi-networks-ltd/, May 2017.

[26] Marcin Nawrocki, Matthias Wählisch, Thomas C. Schmidt, Christian Keil, and Jochen Schönfelder. A Survey on Honeypot Software and Data Analysis. Technical Report arXiv:1608.06249, August 2016.

[27] NetLock. 1 Invalid Embedded SCT. https://crt.sh/?id=473172319, 2018.

[28] Carl Nykvist, Linus Sjöström, Josef Gustafsson, and Niklas Carlsson. Server-Side Adoption of Certificate Transparency. In *PAM*. Springer, 2018.

[29] Devon O'Brien. Certificate Transparency Enforcement in Google Chrome. https://groups.google.com/a/chromium.org/forum/#!msg/ct-policy/wHILiYf31DE, 2018.

[30] Craig Partridge and Mark Allman. Ethical Considerations in Network Measurement Papers. *Communications of the ACM*, 2016.

[31] Vern Paxson. Bro: A System for Detecting Network Intruders in Real-time. *Computer Networks*, 1999.

[32] Carlos Perez. https://github.com/darkoperator/dnsrecon, 2018.

[33] Rapid 7. Forward DNS. https://opendata.rapid7.com/sonar.fdns_v2/, 2018.

[34] The Rook. https://github.com/TheRook/subbrute, 2018.

[35] Quirin Scheitle, Taejoong Chung, Jens Hiller, Oliver Gasser, Johannes Naab, Roland van Rijswijk-Deij, Oliver Hohlfeld, Ralph Holz, Dave Choffnes, Alan Mislove, and Georg Carle. A First Look at Certification Authority Authorization (CAA). *ACM SIGCOMM Computer Communications Review (CCR)*, April 2018.

[36] Quirin Scheitle, Oliver Gasser, Minoo Rouhi, and Georg Carle. Large-Scale Classification of IPv6-IPv4 Siblings with Variable Clock Skew. In *TMA*, June 2017.

[37] Quirin Scheitle, Matthias Wählisch, Oliver Gasser, Thomas C. Schmidt, and Georg Carle. Towards an Ecosystem for Reproducible Research in Computer Networking. In *ACM SIGCOMM Reproducibility Workshop*, 2017.

[38] Ryan Sleevi. Announcement: Requiring Certificate Transparency in 2017. https://groups.google.com/a/chromium.org/forum/#!msg/ct-policy/78N3SMcqUGw, 2016.

[39] Ryan Sleevi. Certificate Transparency in Chrome - Change to Enforcement Date. https://groups.google.com/a/chromium.org/forum/#!msg/ct-policy/sz_3W_xKBNY, 2017.

[40] TeliaSonera. Invalid Embedded SCT. https://crt.sh/?id=295064943, 2018.

[41] Benjamin VanderSloot, Johanna Amann, Matthew Bernhard, Zakir Durumeric, Michael Bailey, and J Alex Halderman. Towards a Complete View of the Certificate Ecosystem. In *IMC*. ACM, 2016.

[42] Matthias Wachs, Quirin Scheitle, and Georg Carle. Push Away Your Privacy: Precise User Tracking Based on TLS Client Certificate Authentication. In *TMA*, 2017.